\documentclass[twocolumn]{aastex631}

\usepackage{xspace}
\newcommand{\hinvMpc}{\,h^{-1}\, {\rm Mpc}\,}

\newcommand{\hMpcinv}{\,h\, {\rm Mpc}^{-1}\,}
\newcommand{\hinvGpc}{\,h^{-1}\, {\rm Gpc}\,}

\newcommand{\Mpch}{\, h^{-1}\mathrm{Mpc}\, }

\newcommand{\kpch}{\, h^{-1}\mathrm{kpc}\, }
\newcommand{\kmax}{\, k_{\rm max}\, }
\newcommand{\knl}{\, k_{\rm NL}\, }

\newcommand{\Lya}{Ly-$\alpha$\xspace}

\newcommand{\qvec}{\mathbf{q}}

\newcommand{\kvec}{\mathbf{k}}

\newcommand{\kpar}{k_{\parallel}}

\newcommand{\kvperp}{\mathbf{k}_{\perp}}

\newcommand{\hacc}{{\small HACC}}
\shorttitle{A Gpc-Scale Hydrodynamic Volume Reconstruction}
\shortauthors{Jacobus et al.} 

\graphicspath{{./}{}}
\usepackage{amsmath}
\usepackage[cal=Euler]{mathalfa}
\usepackage{listings}
\usepackage{bigints}
\usepackage{lipsum}

\setlipsum{%
  par-before = \begingroup\color{gray},
  par-after = \endgroup
}

\usepackage{booktabs}

\begin{document}

\title{A Gigaparsec-Scale Hydrodynamic Volume Reconstructed with Deep Learning}

\author{Cooper Jacobus}
\affiliation{Lawrence Berkeley National Laboratory, Berkeley, CA 94720, USA}

\author{Roger de Belsunce}
\affiliation{Lawrence Berkeley National Laboratory, Berkeley, CA 94720, USA}
\affiliation{Berkeley Center for Cosmological Physics, Department of Physics, University of California, Berkeley, CA 94720, USA}

\author{Sol{\` e}ne Chabanier}
\affiliation{Lawrence Berkeley National Laboratory, Berkeley, CA 94720, USA}

\author{Peter Harrington}
\affiliation{Lawrence Berkeley National Laboratory, Berkeley, CA 94720, USA}

\author{J.D.~Emberson}
\affiliation{Argonne National Laboratory, Lemont, IL 60439, USA}

\author{Zarija Luki\'c}
\affiliation{Lawrence Berkeley National Laboratory, Berkeley, CA 94720, USA}

\author{Salman Habib}
\affiliation{Argonne National Laboratory, Lemont, IL 60439, USA}

\begin{abstract}
The next generation of spectroscopic surveys will map the large-scale structure of the Universe at high redshifts ($2 \leq z \leq 5$) using millions of quasar spectra, enabling major advances in constraining both the standard cosmological model and its extensions. Robust cosmological analyses of these datasets require numerical simulations that both cover Gigaparsec volumes and resolve features on $\sim$10 kiloparsec scales and smaller. However, running such large-volume, high-resolution hydrodynamic simulations is computationally prohibitive. We present a generative deep-learning model that enhances a low-resolution, Gigaparsec-scale ($960 \hinvMpc$) hydrodynamic simulation using a smaller ($80 \hinvMpc$) high-resolution input hydrodynamic simulation  as training data. The resulting enhanced simulation reproduces the line-of-sight  power spectrum to within $\sim 10\%$ and the three-dimensional power spectrum at the $\sim 20\%$ level at intermediate to small scales ($k \lesssim 2 \hMpcinv$).  Our method shows strong promise for producing realistic simulations for cosmological analyses with current surveys such as the Dark Energy Spectroscopic Instrument (DESI) and upcoming next-generation experiments, but further improvements are needed to accurately recover the large-scale modes. We publicly release the enhanced hydrodynamic simulation, along with a halo catalog from a companion $N$-body dark matter simulation to support the calibration of data analysis pipelines for these large-scale surveys.
\end{abstract}


\section{Introduction} \label{sec:intro}
The currently observing Dark Energy Spectroscopic Instrument (DESI) and its planned successors DESI-II and Spec-S5 \citep{Schlegel:22_spec_roadmap,Besuner25:spec_s5} will observe beyond a million Lyman-$\alpha$ (\Lya) forest spectra requiring mocks that approximately span a volume of $(10\hinvGpc)^3$ \citep{Farr:2019xij}, enabling three–dimensional clustering measurements and baryon–acoustic–oscillation (BAO) studies at high redshift $z\gtrsim2$ with sub-percent–level precision~\citep{DESI2016, DESI_BAO_2024, DESI_lya_2024}. 
Extracting cosmological information from these surveys requires creating mock skies that simultaneously (i) cover the full survey footprint to capture large–scale modes and (ii) resolve small-scale features (i.e.,~gas fluctuations at the $\sim20\,\mathrm{kpc}/h$ scale and below) influencing observed summary statistics such as the power spectrum~\citep{Lukic2014, Walther2021, Doughty2023}.  Conventional \Lya hydrodynamic simulations commonly meet the second requirement but at prohibitive computational and memory costs for modeling large volumes.

The \Lya forest, absorption features in the spectra of distant quasars, traces density fluctuations in the neutral hydrogen along the line-of-sight. Observing and correlating a large number of sight lines traces the large-scale structure of our Universe down to $\hinvMpc$ scales and below \citep{McDonald:1999dt, McDonald:2001}. 
Its one–dimensional power spectrum (P1D) already provides leading constraints on neutrino masses and dark–matter models~\citep{Croft1999, McDonald2005, Seljak2005, PDB2015b, Viel2013, Ivanov:2024jtl}. The two-point correlation function provides a unique measurement of the BAO at high redshift \citep{DESI_lya_2024}. DESI is pushing this further by increasing the amount of observed spectra and by  including information beyond the BAO feature either through compressed statistics such as redshift-space distortions (RSD) and the Alcock-Pacy\'nski effect (AP; \cite{Cuceu:2021hlk}) or by using the broadband shape (\emph{i.e.} the full-shape) of the correlations \citep{Gerardi:2022ncj}. Recent advances in measuring the three–dimensional power spectrum (P3D; see, e.g.,~\cite{deBelsunce:2024knf}) and in theoretical modeling using the effective field theory of large-scale structure (EFT; \citealt{McDonald:2009dh, Baumann:2010tm, Carrasco:2012cv}) extended to the \Lya forest \citep{Ivanov:2023yla} allow for a consistent model spanning large to intermediate scales to jointly extract the BAO information as well as the shape of the power spectrum (or correlation function) \citep{deBelsunce:2024rvv, Chudaykin:2025gsh, Hadzhiyska:2025cvk}.  Realizing the full potential of this data set hinges on mock catalogs that reproduce both the P1D and P3D across the $k\sim0.001\hbox{--}10\,h\,\mathrm{Mpc^{-1}}$ range.

State–of–the–art hydrodynamic solvers such as \texttt{Nyx} code \citep{Almgren_2013, Sexton2021}, consistently capture non-linear IGM physics in the context of structure formation \citep{Cen1994}, but  quickly become computationally prohibitively expensive.\footnote{
See also: Illustris TNG~\citep{2017MNRAS.465.3291W, Pillepich:2017jle}, ACCEL$^2$ \citep{Chabanier:2024knr}, CAMELS \citep{CAMELS_presentation, CAMELS_DR1}, PRIYA \citep{Bird:2023evb}.}
Scaling them naively to survey volumes is infeasible: the numbers of grid cells and time steps grow as the scale cubed, and even storing a single snapshot of a $1\,\mathrm{Gpc}/h$ high–resolution run would require ${\sim}10^{15}$ grid elements, or ${>}1\,\mathrm{PB}$ of memory.  Conversely, purely $N$–body or coarse–grid hydrodynamical runs require far less memory but smear out small–scale structure, see, e.g.,~$\textsc{AbacusSummit}$ $N$-body simulations which paint the forest on top of the dark matter field \citep{Hadzhiyska:2023, Hadzhiyska:2025cvk} adopting a fluctuating Gunn-Peterson approximation (FGPA) which has been calibrated on hydrodynamic simulations \citep{Croft98,2022ApJ...930..109Q}. Further, the \Lya Mass Association Scheme (LyMAS; \cite{Peirani:2014,Peirani:2022}), the Iteratively Matched Statistics method (IMS; \cite{Sorini:2016}), Hydro-BAM \citep{2022ApJ...927..230S}, and cosmic-web-dependent FGPA \citep{2024A&A...682A..21S} have been used to connect observed flux to the underlying matter field. Recently, \cite{deBelsunce:2025bqc} presented a perturbative technique to model the \Lya forest directly at the field-level.

\begin{figure*}
  \centering
  \includegraphics[width=\textwidth]{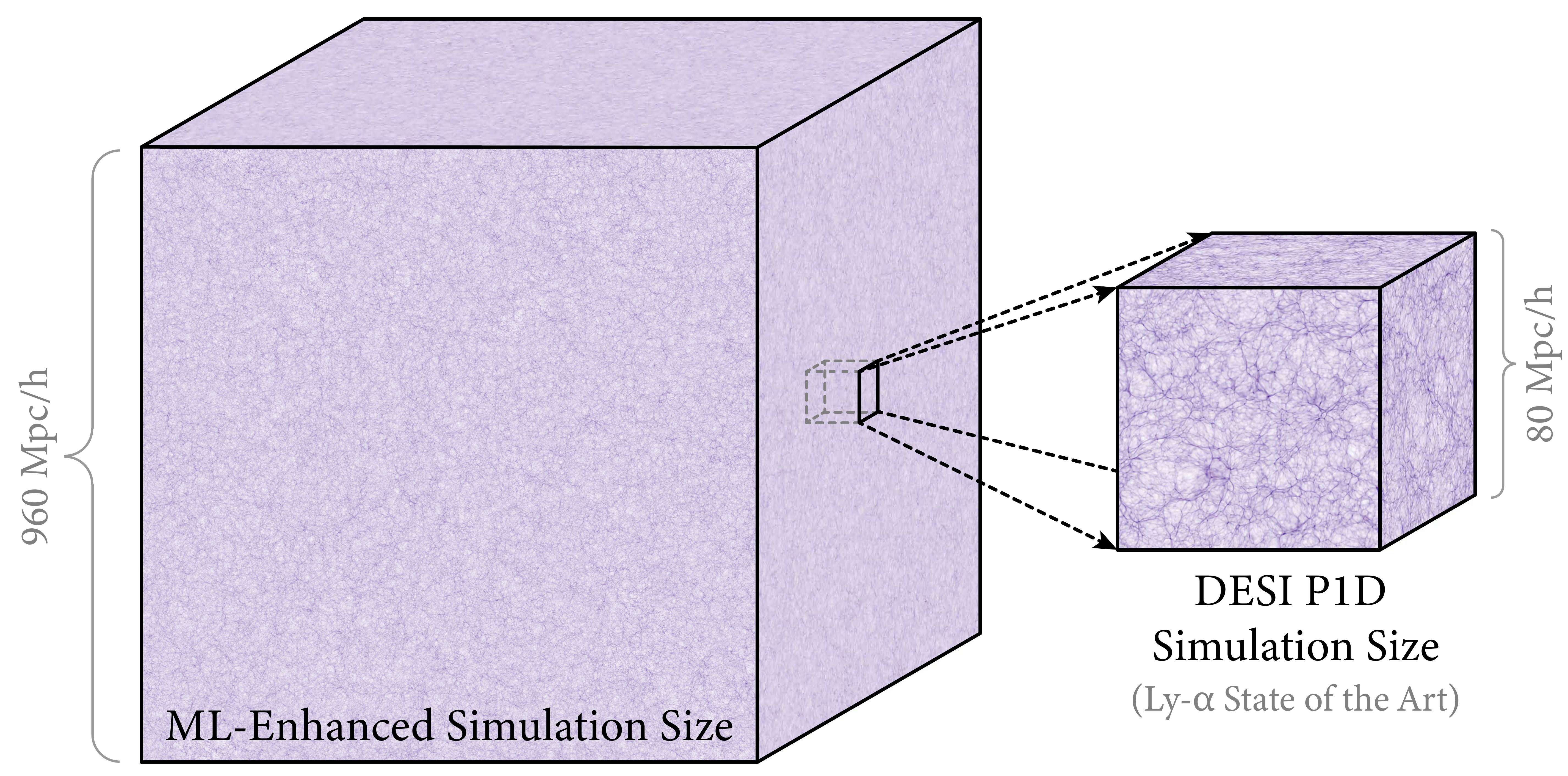}
  \caption{Visualization of the $960 \hinvMpc$ wide baryon density volume we have reconstructed using our machine learning model. The dashed box highlights a region $80 \hinvMpc$ wide, which is the size of our training data and comparable to existing simulations used for DESI \citep{Walther_2021, Walther:2024tcj} and other surveys.}
  \label{fig:big_boy}
\end{figure*}

Simulating the large-scale structure of the universe while simultaneously capturing the relevant small-scale feature at high fidelity is challenging and requires accurate modeling of nonlinear processes in the IGM, such as small-scale gas flows, thermal feedback, and the interplay between gravity and hydrodynamics. To address this critical challenge, we build up on previous work from~\citet{Jacobus_2023} and demonstrate a promising path forward: combining a relatively inexpensive, coarse $960\hinvMpc$ hydrodynamic run with a generative deep‐learning model which we trained to reconstruct the small-scale baryonic physics. Our conditional generative–adversarial network ~\citep{Goodfellow2014,mirza2014conditional} learns to reconstruct small–scale power and correct thermodynamic biases. The result is a $960\,\mathrm{Mpc}/h$ \Lya volume, shown in Fig.~\ref{fig:big_boy}, whose P1D agrees with a fully resolved reference simulation to better than $10\%$ across $k=0.1\hbox{-}5\hMpcinv$, and whose flux PDF matches the reference to ${\sim}10\%$ over most of its dynamic range. The  P3D of the reconstructed box agrees at the $\sim 10-20\%$ level in the range $k\leq 2\hMpcinv$.

The reconstructed simulation we present here spans three orders  of magnitude more volume than the input hydrodynamic \Lya simulations while retaining effective resolution comparable to a simulation on a ${\sim}20\,\mathrm{kpc}/h$ Eulerian grid. We quantify the performance of our model and the accuracy of our mock volume using the one-point probability density function, the P1D and P3D and fit to the latter a perturbative model to measure the bias parameters. The reconstructed hydrodynamic volume is released to the community for survey analysis pipeline calibration and theoretical modeling. While our model is imperfect on large scales, these advances suggest a promising road map for developing survey-scale, high-fidelity mocks skies that are required to interpret the unprecedented influx of data at high redshift from current and future surveys.  In the future work, we plan to address these imperfections using a hybrid approach which combines the current machine learning method with the perturbation theory.


This paper is organized as follows: We first discuss the simulations and data processing techniques we use to train our network in Section \ref{sec:simulations}. We then briefly outline the design of our deep-learning algorithm and the methods used to train it in section \ref{sec:model}. We introduce the theoretical model to validate the reconstruction in Section \ref{sec:EFT_theory}. Our results are presented in Section \ref{sec:results}, where we discuss the accuracy of our algorithm, comment on some of the statistical features of our new, higher-fidelity, Gpc-scale volume, and explore how this volume can be employed for precision cosmology. 
We conclude and discuss future work in Section \ref{sec:conclusion}.

\newpage

\section{Training Data} \label{sec:simulations}

We assemble our training and testing data from pairs of simulations run by the publicly available \texttt{Nyx} code \citep{Almgren_2013, Sexton2021}. The code follows the evolution of dark matter using collisionless self-gravitating Lagrangian particles which are coupled to baryonic ideal gas modeled on a uniform Cartesian grid. Although \texttt{Nyx} is capable of employing adaptive mesh refinement (AMR), we do not make use of this functionality here as the \Lya signal spans the vast low-density volumes and is fully absorbed close to halos where AMR would be relevant. To model the Lyman $\alpha$ forest, \texttt{Nyx} models primordial gas composition of neutral and ionized Hydrogen and Helium, as well as free electrons. This type of simulation is very common in studies of the intergalactic medium and is used as a forward model in virtually any recent inference work using the Lyman alpha power spectrum\citep{Boera2019, Walther2019, PDB2020, Rogers2020, Walther_2021}. 

The simulations are initialized at $z = 200$, using the Zel’dovich approximation~\citep{zeldovich1970}. Transfer functions were generated with the Boltzmann solver code CLASS~\citep{class2011}. The cosmological parameters are set according to the Planck-2016~\citep{Planck2015} model: $\Omega_b = 0.0487$, $\Omega_m = 0.31$, $H_0 = 67.5 \, \mathrm{km}/s/\mathrm{Mpc}$, $n_s = 0.96$ and $\sigma_8 = 0.83$.  For our training data, we produce two pairs of simulations 80 $h^{-1}$Mpc on a side, where each pair has a high- and low-resolution run initialized from the same random initial conditions. The high-resolution simulations have 4096$^3$ elements (both dark mater particles and hydrodynamic grid cells) and are thus much more nearly converged and percent-level accurate, while the low-resolution simulations are much coarser with only 512$^3$ elements. All simulation runs have identical cosmology, and UV background, and the two pairs of high- and low-resolution simulations differ only in the choice of initial condition.  This enables the model to be trained on one pair of simulations and tested on another. We use the software suite \texttt{gimlet} \citep{Friesen2016} to derive the optical depth and flux fields from our hydrodynamic fields and to calculate the relevant summary statistics from these fields.

\begin{figure}[t]
  \centering
    \includegraphics[width=\columnwidth]{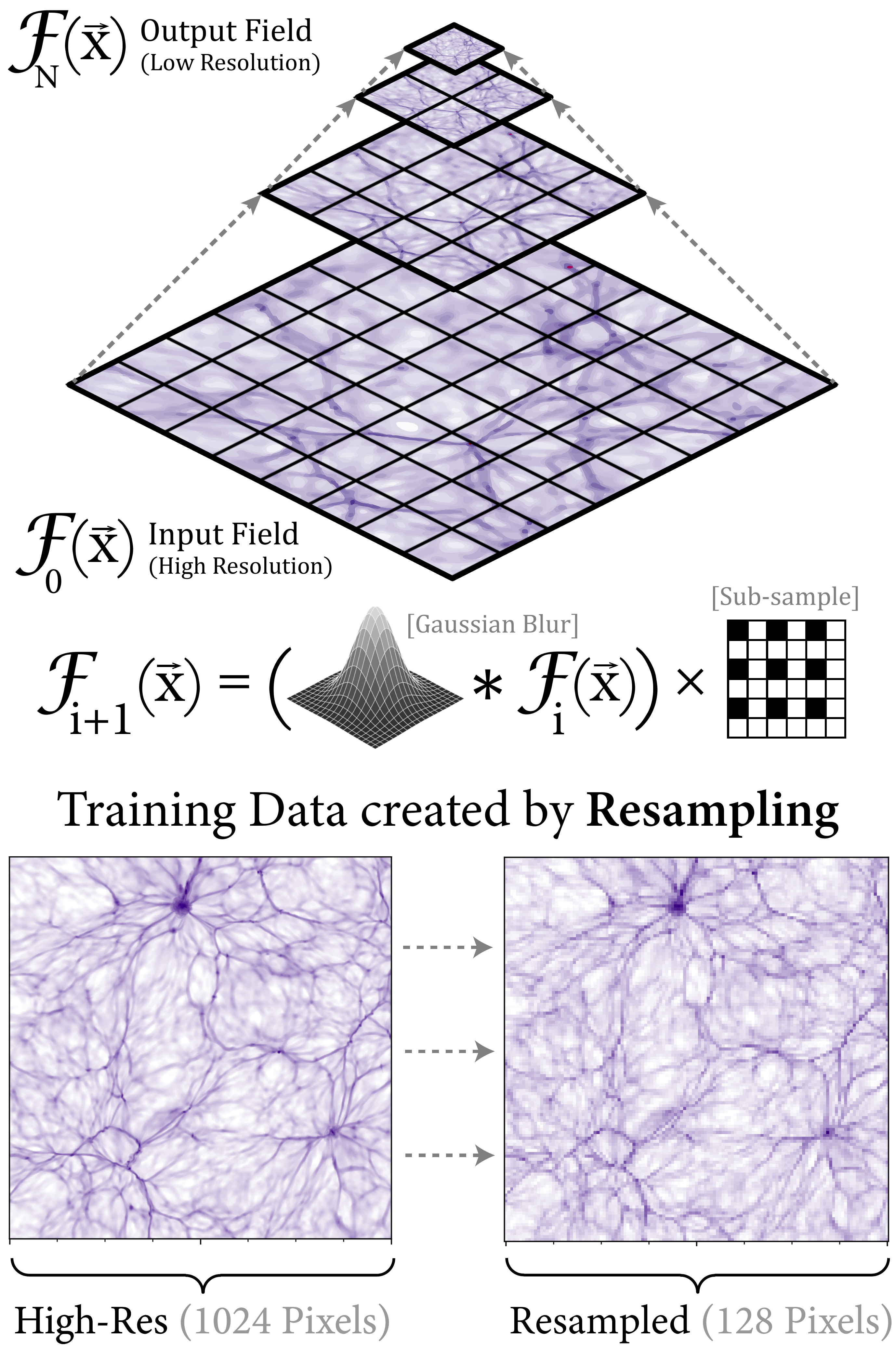}
  \caption{Illustration of the process for creating training data target fields from high-resolution simulations. Fields are iteratively Gaussian blurred and sub-sampled until they match the grid size (and data footprint) of the low-resolution simulations. This method preserves most fine structures while dramatically reducing the memory necessary to store the density fields on our computer. }
  \label{fig:Pyramid}
\end{figure}

In order to generate training data of high spectral fidelity but with a resolution equal to our inexpensive simulations, we use an iterative blurring and resampling method known as a Gaussian pyramid \citep{Pyramid}. We wish to reduce the data footprint of our high-resolution simulations by a factor of 512 while preserving the field's spectral information. We achieve this by iteratively applying a Gaussian blur and subsampling our fields by a factor of two, as depicted in figure \ref{fig:Pyramid}. We describe this method in more detail and demonstrate that it prevents aliasing in \cite{Jacobus_2023}.

\newpage

\section{Machine Learning Model Design}\label{sec:model}

To achieve the science goals described above, we must design a deep learning model to reconstruct physically realistic hydrodynamic fields from coarse-grid simulations, while keeping the memory overall footprint constant. This presents significant challenges due to sharp discontinuities between the input and target fields and inherent uncertainty in the mapping due to the sizable information content gap between resolutions. To address these challenges, we employ a conditional GAN framework \citep{karras2019stylebased, karras2020analyzing} inspired by the fully convolutional TSIT model \citep{TSIT}. This architecture employs a multi-scale encoder-generator structure, incorporating residual blocks to extract and process features at multiple spatial scales. 

Our generator integrates multi-scale Gaussian noise injection, inspired by StyleGAN \citep{karras2019stylebased}, which treats noise injection scales as learnable parameters. This approach enhances stochasticity and better captures physically plausible variability across spatial scales (overcoming the limitations of single-point noise injection). This non-deterministic architecture enables the production of diverse yet realistic ensembles, facilitating the quantification of uncertainty. Point-wise variance across these ensembles of outputs strongly correlates with predictive error, as shown in \citep{Jacobus_2023}.  This novel feature of our model naturally yields an uncertainty estimate that can be propagated throughout our analysis pipeline.

Since the baryon density and temperature fields in the simulations produced by \texttt{Nyx} contain structures spanning several orders of magnitude, their features (such as galaxy clusters and thermal shocks) are most effectively interpreted in logarithmic space. To accommodate this, we pre-process these hydrodynamic fields by log-normalizing them to fit within the range [-1,1] before feeding them into our machine-learning model. 


To ensure the model learns to replicate the statistical properties of the data accurately, we employ a composite loss function composed of several terms. The primary loss component is the point-wise $\mathcal{L}_1$ distance, which stabilizes training and ensures consistency of large-scale features. Additionally, a multi-scale adversarial loss \citep{wang2018pix2pixHD,park2019semantic, TSIT} refines small-scale reconstruction using multiple patch-based discriminators. Given the crucial role of spectral accuracy in our application, we also incorporate a spectral loss term, $\mathcal{L} _ {\textrm{fft}}$, which captures the differences in the truncated Fourier transforms between the predicted and target fields. For a more detailed description of our model architecture, training strategies, and inference methodology, see \citep{Jacobus_2023}.

\section{Theoretical Model} \label{sec:EFT_theory}
To validate the large-volume, low-resolution hydrodynamic simulation that we reconstructed using a generative deep learning model, we fit the resulting three-dimensional power spectrum using a theoretical model based on the effective field theory of large-scale structure (hereafter EFT). The EFT formalism uses a perturbative expansion of the large-scale dynamics by only using the symmetries relevant to the tracer~\citep{McDonald:2009dh,Baumann:2010tm,Carrasco:2013mua,Ivanov:2022mrd} which for the \Lya\ forest are the equivalence principle and rotational invariance around the line-of-sight direction $\hat{z}$, corresponding to the $SO(2)$ group~\citep{McDonald:1999dt,Givans:2020sez,Desjacques:2018pfv,Chen:2021rnb,Ivanov:2023yla,Ivanov:2024jtl, deBelsunce:2024rvv, Belsunce_Sullivan_skewspectrum_2025}. In the following, we briefly summarize the one-loop \Lya EFT model and refer the reader to \cite{Desjacques:2018pfv,Ivanov:2023yla,Ivanov:2024jtl, deBelsunce:2024rvv} for a more complete presentation.  

In linear theory the \Lya flux decrements are modeled using a bias expansion \citep{McDonald:1999dt,McDonald:2001} 
\begin{equation}
\delta_F = b_1\delta + b_{\eta}\eta \,, 
\end{equation}
with the dimensionless gradient of the peculiar velocity, $\eta = -\partial_{\parallel}v_\parallel/\mathcal{H}$ where $\mathcal{H}=aH$ is the conformal Hubble parameter given by the scale factor $a$ and the Hubble parameter $H$. The resulting tree-level power spectrum is connected to the linear matter power spectrum via
\begin{equation}
P^{\rm tree}(k,\mu) = (b_1 + b_{\eta}f\mu^2)^{2} P_{\rm lin}(k)\,,
\end{equation}
where $k$ is the Fourier wavenumber, $\mu$ the angle of $k=\{\kpar,\kvperp\}$ to the line-of-sight, $\mu \equiv \kpar/k$, $f$ is the (linear) growth rate, and  the two linear bias parameters are $b_1$ and $b_{\eta}$. All quantities are evaluated at the redshift of the simulation which we suppress in the following. 

Schematically, the one-loop EFT power spectrum is based on four components
\begin{equation} \label{eq:Pmodel}
    P^{\rm theory} = P^{\rm tree} + P^{\rm 1-loop} + P^{\rm ct} + P^{\rm st}\,,
\end{equation}
where $P^{\rm tree}$ is the infrared resummed linear theory power spectrum.\footnote{This is computed using the Boltzmann solver \texttt{CLASS-PT}~\citep{Diego_Blas_2011,Chudaykin:2020aoj}.}
The one-loop \Lya power spectrum component is given by 
\begin{align}
P^{\rm 1-loop}& (k,\mu) 
=2\int_{\qvec} K_2^2(\qvec,\kvec-\qvec)
P_{\text{lin}}(|\kvec-\qvec|)P_{\text{lin}}(q)  \nonumber\\
&\quad + 6 K_1(\kvec)P_{\rm lin}(k)\int_{\qvec} K_3(\kvec,-\qvec,\qvec)P_{\text{lin}}(q)\,,
\end{align} 
with higher order redshift-space kernels, $K_{2,3}$ given in Eq.~(3.19) in \cite{Ivanov:2023yla} with the short hand notation $\int_{\qvec}\equiv\int \frac{d^3q}{(2\pi)^3}$ to denote the three-dimensional integral over $\qvec$. The counter terms scale as $k^2P_{\rm lin}(k)$ 
\begin{align}
\label{eq:shoch}
P^{\rm ct}(k,\mu) =
&-2(c_0+c_2\mu^2+c_4\mu^4)K_1(\kvec)k^2 P_{\rm lin}(k)\,,
\end{align} 
with the parameters $c_{0,2,4}$ controlling the angular dependence and the stochastic contributions are a function of a constant shot noise piece and a scale-and angle-dependent term, accounting for small-scale physics\footnote{We emphasize that the EFT is only valid at larger scales than the non-linear scale, $\knl$, and should be treated as phenomenological beyond it. The scale is defined as $\Delta^2 = k^3P_{\rm lin}(k,z)/(2\pi^2)\sim 1$, i.e., where the dimensionless power spectrum reaches unity. For the current simulations this corresponds to $k_{\rm NL} \approx 7 \hMpcinv$ at the redshift $z=3.0$.}
\begin{align}
P^{\rm st}(k,\mu) =
&P_{\text{shot}}+a_0\frac{k^2}{\knl^2}+a_2\frac{k^2\mu^2}{\knl^2}\,.
\end{align}
Measurements of these parameters on different \Lya simulations have been presented in \cite{Ivanov:2023yla, deBelsunce:2024rvv, Hadzhiyska:2025cvk}.

The one-loop \Lya power spectrum is described by the following set of nuisance parameters 
\begin{equation}
\label{eq:nuissance_param} \{b_1, b_\eta, b_2, b_{\mathcal{G}_2}, b_{(KK)_\parallel}, b_{\Pi^{[2]}_\parallel}, b_{\delta \eta}, b_{\eta^2}\}\,,
\end{equation}
in addition to the cubic EFT terms 
\begin{equation}
\label{eq:cubic_param} \{b_{\Pi^{[3]}_\parallel},b_{(K\Pi^{[2]})_\parallel},b_{\delta\Pi^{[2]}_\parallel},b_{\eta\Pi^{[2]}_\parallel}\}\,,
\end{equation}
and the counter and stochastic terms 
\begin{equation}
\label{eq:ct_param}  \{c_{0,2,4},a_{0,2},P_{\rm shot}\}\,,
\end{equation}
yielding a total of 18 parameters. We adopt the prior choices of table 1 in \citet{deBelsunce:2024rvv} and analytically marginalize over the parameters in Eqs.~\eqref{eq:cubic_param}-\eqref{eq:ct_param}. 

To fit the 3D power spectra, we follow the standard procedure in \cite{Givans:2022qgb} and sample the $\chi^2$ function (jointly fitting all four angular and all $k$-bins)
\begin{equation} \label{eq:chi2}
    \chi^2 = \sum_i \frac{\left[P_i^{\rm data}-P^{\rm model}_i\right]^2}{2 \left(P_i^{\rm data}\right)^2/N_i}\,.
\end{equation}
The measured power spectrum is denoted by $P_i^{\rm data}$ in $k$ and $\mu$-bins  with $N_i$ Fourier modes per bin, $P^{\rm model}_i$ is the EFT theory prediction. The error bars are computed assuming a Gaussian covariance, given by $P^{\rm data}_i\sqrt{2/N_i}$, which assumes the measured power along each of the axes to be independent. Note that we account for mode discreteness by evaluating the model vector at the power spectrum-weighted mean $k$ and $\mu$ for each bin, computed via $\overline{x}=\sum_i P_i^{\rm data}x_i/\sum_i P_i^{\rm data}$ for $x\in \{k,\mu\}$.  For the present analysis we do not vary cosmology and use the input cosmology of the simulation and use $\kmax=1 \hMpcinv$ as baseline maximum wavenumber for the large and $\kmax=2 \hMpcinv$ for the smaller volumes.\footnote{For a discussion on the impact of scale cuts and biases induced by higher-order corrections see figure 3 in \cite{deBelsunce:2024rvv}.}

\section{Results} \label{sec:results}

We present a new hydrodynamic simulation volume reconstructed using our deep-learning method. The simulation is a cube 960 $\hinvMpc$ wide represented by $6144^3$ volumetric cells containing values of dark matter density, baryon density, temperature, velocity, and \Lya optical depth. (For reference the high and low-resolution training volumes cover an 80 $\hinvMpc$ wide cube with $4096^3$ and $512^3$ cells respectively, see figure \ref{fig:big_boy}). Our simulation was run using \texttt{Nyx} at a resolution exactly matching that of our low-resolution training data, was likewise initialized at $z = 200$, and with the exact same cosmology and UV background. We take a simulation output at a redshift of $z = 3$ and then pass it through our deep-learning model to improve the small-scale fidelity. The reduced simulation output in single precision requires 11 Terabytes of computer memory, for the grid representations of dark matter and baryon quantities, as well as the \Lya optical depth. A high-resolution simulation of this volume would require $>5.6$ Petabytes.

In the following subsections, we explore the accuracy of our ML-enhanced 960 $\hinvMpc$ wide simulation by examining a few key summary statistics of the hydrodynamic fields and their corresponding \Lya flux fields. We demonstrate the power of this larger volume to contain cosmological models by using it to fit an effective field theory (EFT) model, introduced in Section \ref{sec:EFT_theory}. 


\subsection{Hydrodynamic Fields}
\begin{figure}
  \centering
    \includegraphics[width=\columnwidth]{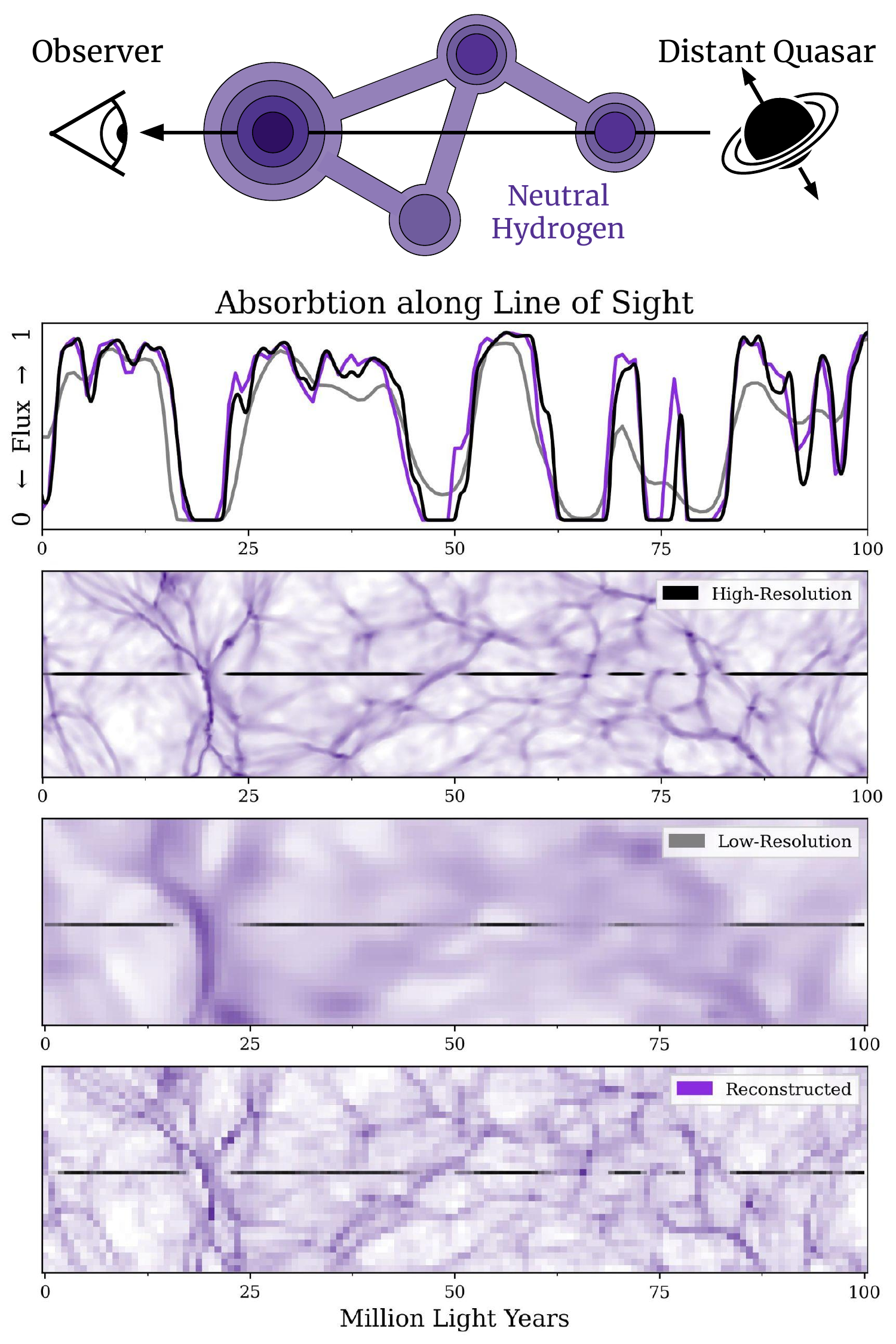}
  \caption{Illustration of the effects of resolution on \Lya absorption features. We plot slices of Baryon Density $\sim 30$ Mpc wide for High-Res, Low-Res, and Reconstructed fields, visualized in log-space.
  Notice that absorption features are spectrally sharpest in the high-resolution simulation but that some spectral detail is recovered by our ML model. }
  \label{fig:Absorbtion_LOS}
\end{figure}

\begin{figure*}[t]
  \centering
  \includegraphics[width=\textwidth]{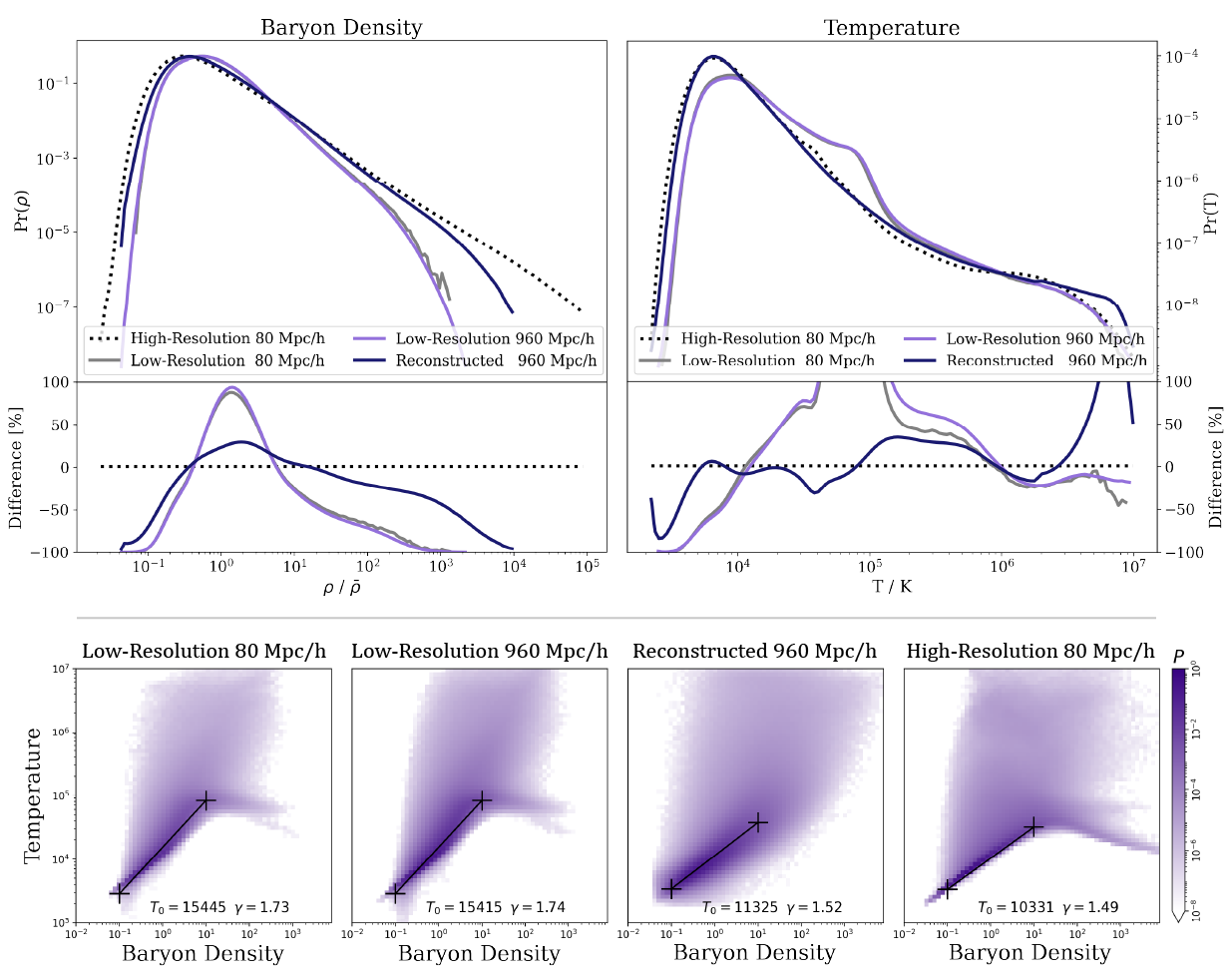}
  \caption{(Top) The probability density functions of the predicted density and temperature fields (dark blue lines) compared against the reference distributions from a target high-resolution simulation (dotted lines). (Bottom) The density–temperature phase distribution of the gas in our model prediction compared against the reference distributions from the low-resolution and high-resolution simulations. For each output we plot the best-fit power-law relationship between density and temperature, our model presents an improvement in the slope of this power-law, $\gamma$, when compared against the low-resolution simulation, but produces a much broader distribution.} 
  \label{fig:2D_hist}
\end{figure*}

The baryon density field of the 960 $\hinvMpc$ wide volume we reconstructed using a generative deep learning model is visualized in figure \ref{fig:big_boy}, the process of creating target and training data in figure \ref{fig:Pyramid} and the improvement is visually demonstrated in figure \ref{fig:Absorbtion_LOS}. We examine the statistical properties of the baryon density and temperature in figure \ref{fig:2D_hist}. We compare the probability distributions of our baryon density and temperature to those from a `true' high-resolution simulation of a much smaller volume. The output of our machine-learning model presents a clear improvement over the low-resolution simulation volume; both distribution peaks have been noticeably corrected, and the distribution shapes have been adjusted to better match that of the high-resolution baseline. 

While the machine learning model does capture the center of the distributions of the baryon density and temperature reasonably well, improving the low-resolution distributions greatly, it fails at accurately capturing the tails of these distributions. This is due to an under-representation of these regimes in the training data by a factor of $10^5-10^6$, and therefore a lack of incentive for the model to accurately reproduce them.


We show the 2D density–temperature distribution in the bottom panel of figure \ref{fig:2D_hist}. While the low-resolution distributions differ from the high-res case, they agree well with each other, indicating that the small volume captures the essential diversity of the large box. Because of our downsampling scheme, much of the fine density–temperature structure is smoothed out, and the condensed, galaxy-forming tail is not well reproduced. However, this regime represents a very small fraction of the total volume. More importantly, our method significantly improves the recovery of the diffuse gas power-law slope, $T = T_0 (\rho / \langle \rho \rangle)^{\gamma-1}$, reducing the error from $\sim17\%$ to $2\%$ as a fractional difference of the high-res slope, $\gamma=1.49$. This improvement strongly affects the \Lya\ signal and provides sharper constraints on the IGM’s ionization history.

\subsection{Lyman-alpha Fields} \label{subsec:lya_fields}

\begin{figure*}[t]
  \centering
  \includegraphics[width=\textwidth]{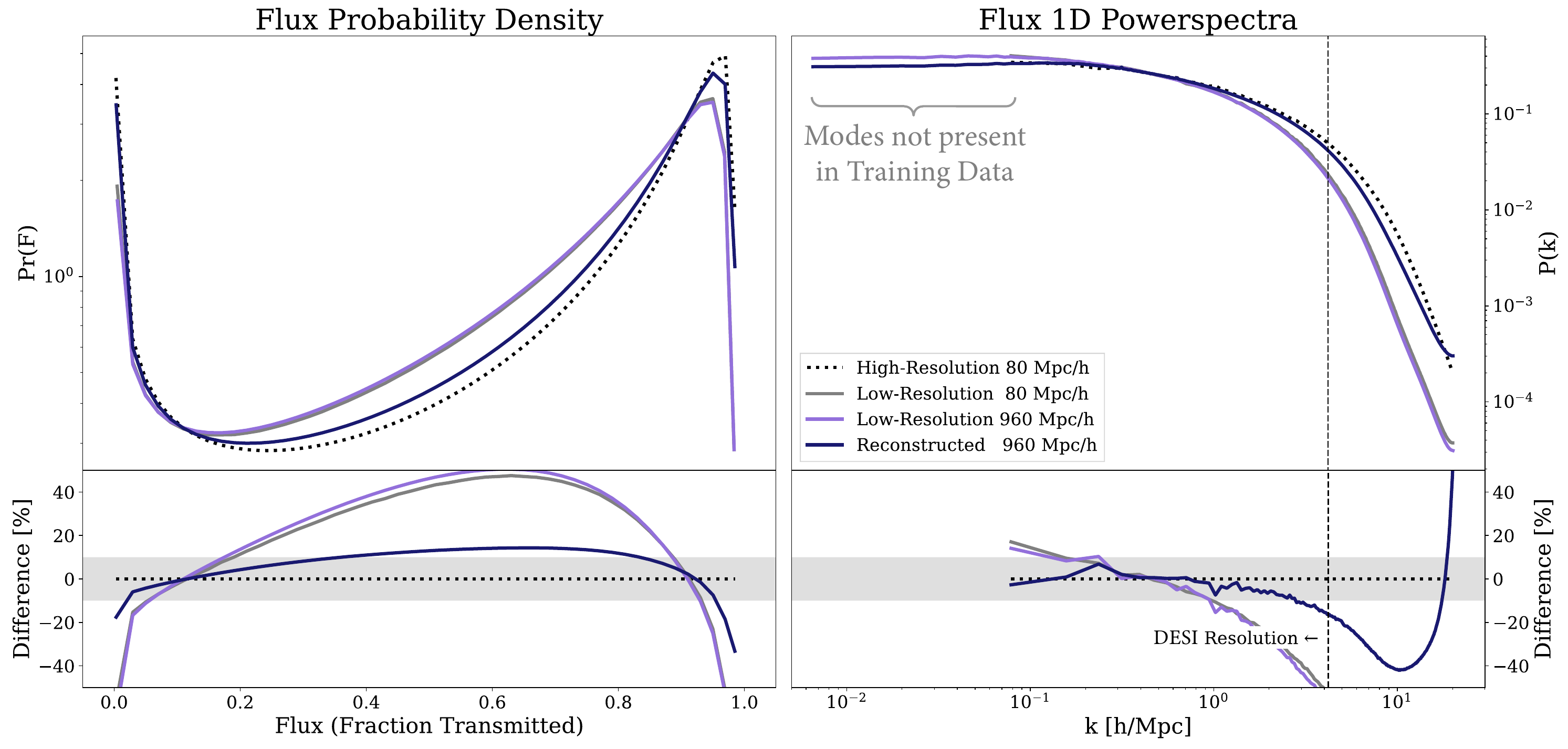}
  \caption{The probability density (PDF, left panel) and 1D power spectrum (P1D, right panel) of \Lya flux in redshift space for four related simulation volumes: the high and low-resolution Nyx simulations, the resized target, and our network output. The grey band in the residual highlights the 10\% error range around the smaller, 80 $\hinvMpc$ high-resolution simulation from our validation dataset. The strong P1D agreement demonstrates that small and intermediate-scale fluctuations are well captured; in contrast, discrepancies in the PDF arise because the model’s limited correlation length prevents reconstruction of correct large-scale phases and spatial correlations. 
  }
  \label{fig:flux_stats}
\end{figure*}

Lyman-$\alpha$ absorption maps are calculated using the \texttt{gimlet} code~\citep{Friesen2016} along lines-of-sight or ``skewers" that pass through the entire volume, parallel to one of the axes, simulating a random distribution of quasar spectra which probe our simulated fields, comparable to the data expected from spectroscopic surveys like DESI. The aggregate of all such lines-of-sight forms the \Lya ``field", representing an absorption map of the entire volume, including redshift space distortions along one of the axes. We calculate some key summary statistics of this field using \texttt{gimlet} and compare them to the target and input simulations to evaluate our method.


We find that our machine learning method reliably reconstructs key \Lya forest statistics across a broad range of scales. Figure 5 compares the flux probability density function (PDF, left) and the 1D power spectrum (P1D, right) between the high-resolution target, the low-resolution baseline, and our ML-enhanced simulation. The ML model reproduces the P1D with high accuracy across all scales, while deviations in the PDF are more pronounced, particularly in the sparse areas of the distribution. This indicates that the method effectively captures the amplitude of small-scale flux fluctuations but is less accurate at recovering the detailed phase information needed to fully reproduce the PDF.


In particular, the flux probability density function (PDF) carries information about the non-linear, halo-scale physics and reflects the thermal phase of the intergalactic medium (IGM), which cannot be accessed through power spectrum measurements alone.
The values of this metric for the input simulations (gray and purple lines in figure \ref{fig:flux_stats} above) are off by as much as 50\% when compared to the high-resolution target for intermediate flux values $(F\sim0.5)$ due to the diffusivity of the low-resolution hydrodynamic fields, which would lead to significant errors if used for cosmological analysis. We show in figure \ref{fig:flux_stats} that the PDF of our reconstructed \Lya field (black line) is within about 10\% of the true (high-res) values. This offers a substantial improvement over the low-resolution input but suggests that the thermal phase of the IGM is still inaccurately captured.

Complementing the information captured by the flux PDF, the 1D power spectrum (P1D) of the \Lya forest offers another key statistical measure that reflects the scale-dependent clustering of transmitted flux along quasar sightlines, constraining both the properties of the IGM and critical cosmological parameters \citep{McDonald2005, PalanqueDelabrouille2013}. The 1D Power Spectrum (P1D) is one of the most important observables that can be predicted by our cosmological models and is especially sensitive to the amplitude and shape of primordial density fluctuations and other large-scale physics. We plot the mean power spectrum $\langle P(k) \rangle$ averaged over modes with magnitude k from all skewers, as done in \cite{Lukic2014}. We show comparisons between the low-resolution, ML-reconstructed, and `true' high-resolution 1D powerspectra in Figure \ref{fig:flux_stats} above. 

Our reconstructed \Lya P1D is in agreement with the high-resolution test volume to within 10\% for wavenumbers smaller than 5 h/Mpc and is in better agreement than the low-resolution simulation for all wavenumbers present in the test volume. At smaller scales, the remaining differences between our reconstructed P1D and the high-resolution target likely arise from imperfect reconstruction of the non-linear density-temperature relationships and thermal broadening effects that influence the detailed shape of absorption features. While our model captures the overall amplitude of small-scale power remarkably well, there are subtle differences in the thermal state of the IGM (see the difference in the tail of the phase distribution in fig \ref{fig:2D_hist}) that affect the fine structure and widths of some individual absorption lines, leading to residual discrepancies in statistics like the P1D that are sensitive to these details. Fortunately, most of these discrepancies impact the power spectrum on scales smaller than what DESI can resolve.

\begin{figure*}[t]
  \centering
  \includegraphics[width=\textwidth]{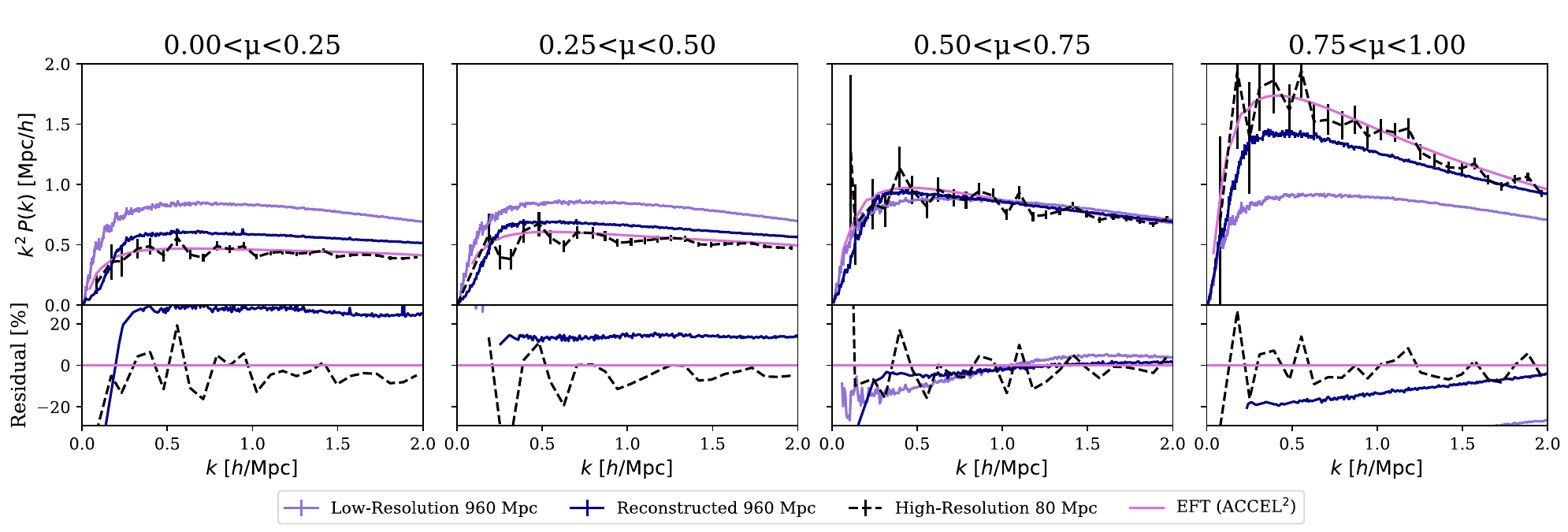}
  \caption{The three-dimensional power spectrum (P3D) of the \Lya forest for four different simulations shown as a function of Fourier wavenumber, $k$, and the cosine of the angle with the line of sight, $\mu = k_{\parallel}/|\bf{k}|$: (i) low-resolution $960 \hinvMpc$ box in purple; (ii) the reconstructed $960 \hinvMpc$ box in blue; (iii) the high-resolution input $80 \hinvMpc$ simulation in black dashed; and (iv) EFT fits to the ACCEL$^2$ simulation in pink with a box of length $L=160 \Mpch$ \citep{Chabanier:2024knr,deBelsunce:2024rvv}. The panels show the 3D power spectrum divided into four separate $\mu$ bins. The generative deep learning model is able to reconstruct the angular evolution for the $960 \hinvMpc$ box and an overall agreement at the $\sim 20\%$ level.}
  \label{fig:p3d}
\end{figure*}

\begin{figure*}[t]
  \centering
  \includegraphics[width=1.0\textwidth]{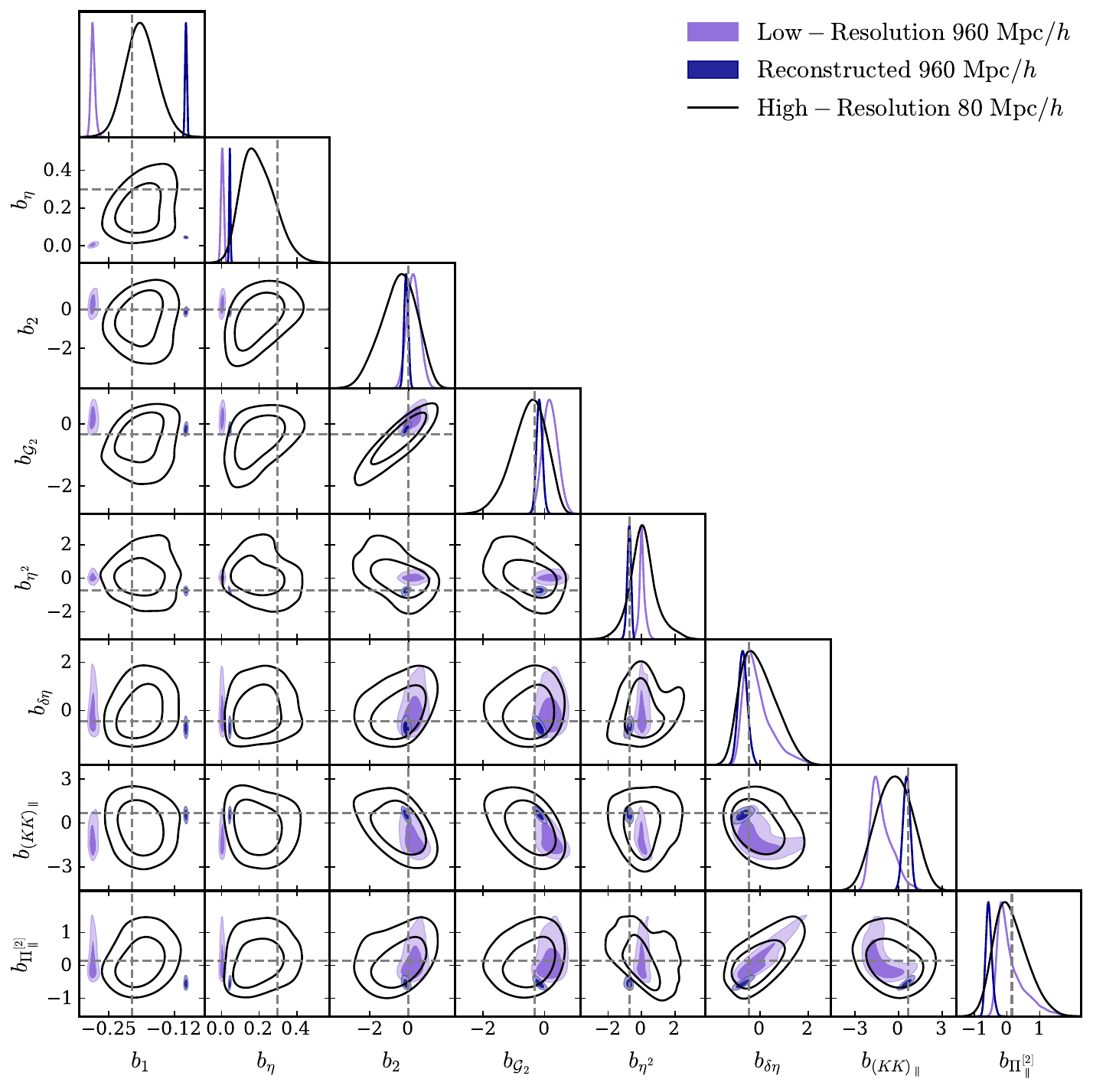}
  \caption{Drift plot of the quadratic bias parameters fit to three simulations: (i) the low-resolution $960 \hinvMpc$ (purple), (ii) the reconstructed 960 $\hinvMpc$ box (blue); and (iii) the input high-resolution 80$\hinvMpc$ simulation (gray). We show the best-fit 1D and 2D marginalized posteriors using a maximum wavenumber of $k_{\rm max}=1 \hMpcinv$ ($k_{\rm max}=2 \hMpcinv$) for the fits of the large (small) boxes.  We find promising agreement in the higher-order EFT bias parameters but note the the linear bias parameters ($b_1$ and $b_\eta$) yield discrepant results at the $2-3\sigma $ level. The quality of the fits is shown in figure \ref{fig:EFT_fits}.}
  \label{fig:corner-plot}
\end{figure*}

\newpage
While our machine-learning model is only able to make adjustments to the field at small scales, by construction, it is worth noticing how the total power at very large scales (ie. low wavenumbers) is affected. Our reconstruction produces a decrease of about 20\% in the power with respect to the low-resolution Gpc-scale input simulation. This is most likely due to the insufficient correlation length (receptive field) of our ML model, which cannot adequately capture very large-scale modes beyond the 80 $\hinvMpc$ training volume size and which may slightly smooth-out some of the very large-scale modes because it was impossible to represent them in the training data. We present some ideas for how to address this in future projects in section \ref{sec:conclusion}.


The further quantify the performance of our model, we present the 3D power spectrum (P3D) of the \Lya flux in figure \ref{fig:p3d}. 
We present the 3D power spectrum as a function of Fourier wavenumber $k$ and the angle between the wave-vector and the line of sight, defined as $\mu \equiv k_{\parallel}/|\bf{k}|$, illustrated in four bins with $\Delta \mu = 0.25$. Larger (smaller) $\mu$ values bin closer to (further from) the line of sight. We compare four power spectra: (i) the low-resolution $960 \hinvMpc$ box in purple; (ii) the reconstructed $960 \hinvMpc$ box in blue; (iii) the high-resolution input $80 \hinvMpc$ simulation in black dashed; and (iv) for comparison we show EFT fits to the ACCEL$^2$ simulation in pink with a box of length $L=160 \Mpch$ with $N=6144^3$ particles, yielding a physical resolution of $25 \kpch$ presented in \cite{deBelsunce:2024rvv}. We emphasize two aspects of this figure: First, the low-resolution simulation does not evolve as a function of $\mu$ as expected from the high-resolution input simulation. For the reconstructed box we see an evolution with angle showing that the generative model is able to inject this information. Second, both high-resolution simulations (which share the same cosmology but use different hydrodynamic prescriptions) agree at the power spectrum level, validating the used training simulation. 
We find that, for all four $\mu$ bins, the power spectrum of the reconstructed simulation achieves a clear improvement, from an error of over 100\% in the case of the low-resolution simulation down to around 20\% with our method. Our ML method mostly recovers the $\mu$-dependence, shifting the spectrum towards the target curve in all $\mu$ bins and at all scales. 


\subsection{Inferring parameters from the simulations} \label{sec:cosmo_EFT}
In this section, we further quantify the performance of the generative model to reconstruct the low-resolution simulation and comment on their applicability for cosmological data analysis. We fit the power spectra shown in Fig.~\ref{fig:p3d} with the EFT theory model introduced in Sec.~\ref{sec:EFT_theory} and compare their bias parameters given in Eqs.~\eqref{eq:nuissance_param}-\eqref{eq:ct_param}. The key question we are interested in answering is how does the reconstruction affect the large and small scales. EFT offers a pathway to explore this question as the linear bias parameters are (mostly) sensitive to the large-scale modes while the higher-order bias parameters are dominated by small-scale modes.

In Fig.~\ref{fig:corner-plot} we show the marginalized 1D and 2D posteriors for the eight bias parameters for the input and reconstructed simulations used in this work. The low-resolution simulation is shown in purple, the reconstructed in blue, and the high-resolution input in gray. Following baseline expectations, the large volume of the coarse and reconstructed hydrodynamic simulation give access to many quasi-linear modes yielding tight constraints on the bias parameters. For the higher-order bias parameters these results are in qualitative agreement with the contour plots presented in \cite{deBelsunce:2024rvv}. The corresponding best-fit EFT spectra are shown in figure \ref{fig:EFT_fits} with the mean best-fit values tabulated in table \ref{tab:EFT_fits}.

It is interesting to note that the higher-order bias parameters from the low- and high-resolution hydrodynamic simulations are fully consistent with each other. The reconstruction sharpens the marginalized posteriors improving the consistency with the input high-resolution simulation. This is in contrast to the linear bias parameters, which are constrained mostly from large-scale modes. The marginalized posteriors on  $b_1$ shows tight constraints with the reconstruction not improving the agreement with the input high-resolution simulation. The velocity bias $b_\eta$ is in (slightly) better agreement with the input simulation. The discrepancy is also visible when assessing the quality of the fits in figure \ref{fig:EFT_fits} for the reconstructed volume. The main reason for the discrepancy in the linear bias parameters  stems from the short correlation length ($\sim 20 \hinvMpc$) of our model. Thus the reconstructed ML box does not have any large- to intermediate-scale information from the input simulation resulting in unrealistic (or even arbitrary) values of $b_1$ and $b_\eta$. This maximum correlation length is determined by the structure of the convolutional neural network, and by the memory limitations of the GPUs used for inference. We emphasize that the reconstruction does improve the small scale information of the coarse large-volume simulation as seen from the agreement of the P1D and the higher-order bias parameters. 





\subsection{Halo Catalog}

In addition to the hydrodynamic simulation discussed above, we performed a companion N-body simulation of the $960 \hinvMpc$ box using the Hardware/Hybrid Accelerated Cosmology Code \citep[\hacc;][]{habib/etal:2016}. The initial noise field of the original \texttt{Nyx} hydrodynamic simulation was used to generate positions and velocities of $6144^3$ dark matter particles at a starting redshift of $z = 200$. The particles were subsequently evolved to $z = 3$ using gravitational forces via typical N-body methods with a Plummer softening length of 1.56 kpc/h. To produce a robust dark matter halo catalog from this n-body simulation, halos were identified using the friends-of-friends (FOF) method from the \texttt{ArborX} library \citep{prokopenko/etal:2024}, which is built into the parallel analysis framework of \hacc. We employ a linking length of $b = 0.168$ and assign the center of each FOF halo as the location of its most bound particle. Next, spherical overdensity (SO) halos are constructed around the center position of each FOF halo using an overdensity criterion of 200 times the critical density. 

We make this halo catalog available alongside our hydrodynamic simulation so that is may be used to calibrate analysis techniques which combine \Lya fields with galaxy/quasar/halo data. The combination of these two data sets will allow cosmologists to explore the rich cross-correlations between the Lyman-$\alpha$ forest and other cosmological tracers, developing new methodologies for extracting physical insights from existing and future surveys.

\section{Conclusion} \label{sec:conclusion}
We present a generative deep-learning model, introduced in section \ref{sec:model}, that enhances a low-resolution, Gigaparsec-scale ($L=960\,h^{-1}\mathrm{Mpc}$) hydrodynamic simulation using a high-resolution ($L=80\,h^{-1}\mathrm{Mpc}$) input hydrodynamic simulation as training data, introduced in section \ref{sec:simulations}. The reconstructed simulation we present here is an essential step towards meeting the demands of current and next-generation spectroscopic surveys, which require both massive survey volumes and accurate small-scale modeling that have previously been out of reach with traditional simulation methods due to computational bottlenecks. By bridging the gap between computational affordability and physical fidelity, an approach like ours can enable the scientific community to create mock universes suitable for precision cosmological analysis, calibration of data pipelines, and testing of new statistical techniques, capabilities that are indispensable for extracting robust constraints from \Lya forest observations. Without such reconstructions, robust interpretation of  \Lya forest data, especially in the context of large-scale structure, neutrino mass, and dark matter studies—would remain severely limited by either computational cost or insufficient resolution. Our method allows us to produce a mock universe over 1000 times larger than an input state-of-the-art simulation.

This volume was reconstructed from a relatively low-resolution simulation, using a deep-learning algorithm that we designed and optimized for this specific purpose. We have demonstrated that this translation method provides an accurate reconstruction over a wide range of scales down to $\sim$ 0.3 $\hinvMpc$
Our method provides a far less computationally expensive way to produce volumes like this one of large scales. Our network can be  trained on a relatively small volume for some given cosmological and astrophysical parameters and then used to reconstruct a (arbitrarily) large volume. 

\newpage
Although our model presents a remarkable improvement in the fidelity and statistical accuracy of the reconstructed \Lya forest, particularly at the smaller scales, which cannot be accurately represented by low-resolution hydrodynamic simulations, there are still some notable shortcomings, particularly on large scales. 
As shown in figures \ref{fig:corner-plot} and \ref{fig:EFT_fits}, the reconstructed \Lya forest does not accurately capture the large-scale modes. Whilst increasing the size of the training set to include multiple high-resolution realizations would further reduce the uncertainty at intermediate scales (which are already very small as shown in the middle panel of figure \ref{fig:EFT_fits}), the largest improvement will stem from increasing the model's limited correlation length it can internally process.



Further, we also note some remaining shortcomings in our model's hydrodynamic field reconstructions, particularly in and around the larger clusters and shocks. While these regions represent a very small percentage of the universe by volume and have only a marginal effect on the \Lya statistics, they are of significant astrophysical interest and affect the hydrodynamic statistics shown in figure \ref{fig:2D_hist}. Because these regimes are especially dependent on subgrid dynamics, they are poorly captured by the input low-resolution simulation and present the greatest challenge to our machine learning model. However, our technique presents a significant improvement over others at capturing these shocks and accurately reproducing the statistics of the \Lya forest.

We have shown that our reconstruction model is able to inject small-scale physics into larger-volume, low-resolution simulations. This is the first step towards constructing high-fidelity, large-volume mocks for surveys like DESI, WEAVE-QSO \citep{2016sf2a.conf..259P}, the Prime Focus Spectrograph \citep[PFS;][]{2022PFSGE}, and 4MOST \citep{2019Msngr.175....3D}. The most pressing shortcoming of the present model is that it cannot accurately reconstruct large-scale modes due to the short correlation length of the deep learning architecture. 
We leave the exploration of a hybrid approach (see, e.g., \cite{Modi:2023drt}) combining the present method for small-scale reconstruction and capturing (and constraining) the large scales using perturbation theory to future work. Recent progress in modeling the \Lya forest directly at the field-level \citep{deBelsunce:2025bqc} offers an interesting pathway for this hybrid approach.

\newpage
\section*{Data Availability} \label{sec:Data}
The complete machine learning–enhanced simulation volume, including all hydrodynamic fields, Lyman-alpha optical depth fields, and the associated dark matter halo catalog are publicly available for access and download from the National Energy Research Scientific Computing Center (NERSC) `Science Gateways' portal: 

\begin{center}
\href{https://portal.nersc.gov/project/nyx/Gpc/}{portal.nersc.gov/project/nyx/Gpc/}
\end{center}

We invite members of the community to access these resources to experiment with this novel data-product and conduct further analysis to enable future large-scale cosmological studies.

\vspace{20pt}
\section*{Acknowledgments} \label{sec:Acknowledgments}
This work was supported by the Department of Energy, Laboratory Directed Research and Development program (PI: Zarija Luki\'c) at Lawrence Berkeley National Laboratory. Production of auxiliary simulations was supported by the DOE’s Office of Advanced Scientific Computing Research and Office of High Energy Physics through the Scientific Discovery through Advanced Computing (SciDAC) program. This research used resources of the National Energy Research Scientific Computing Center, a DOE Office of Science User Facility supported by the Office of Science of the U.S. Department of Energy under Contract No. DEC02-05CH11231. This research also utilized the resources of the Argonne Leadership Computing Facility, which is a DOE Office of Science User Facility supported under Contract DE-AC02-06CH11357.

C.J.~gratefully acknowledges scholarship support from the Goldwater Foundation and the NASA Astronaut Scholarship Foundation.

\newpage
\bibliographystyle{aasjournal}
\bibliography{main}

\appendix

\section{Fit details}
\begin{figure}[h]
    \centering
    \includegraphics[width=0.32\linewidth]{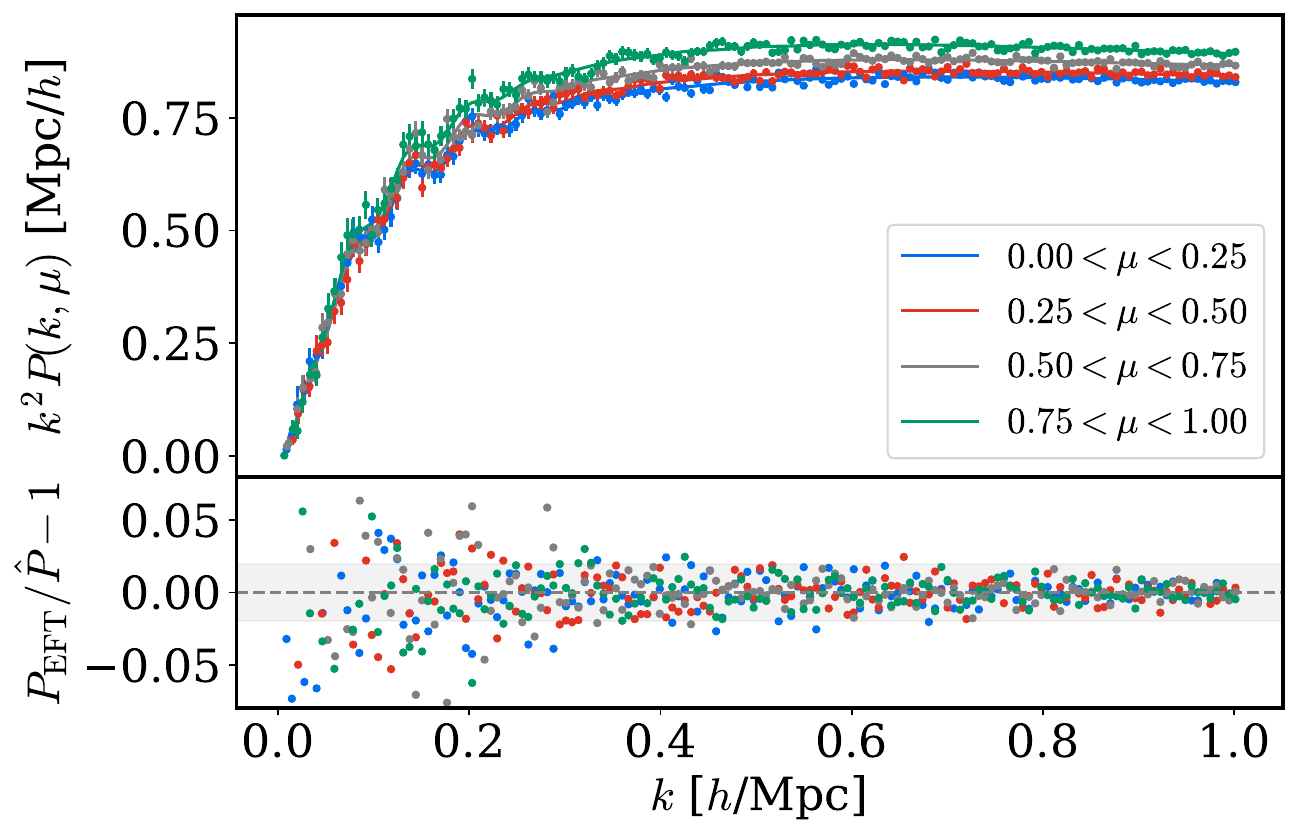}\hfill
    \includegraphics[width=0.32\linewidth]{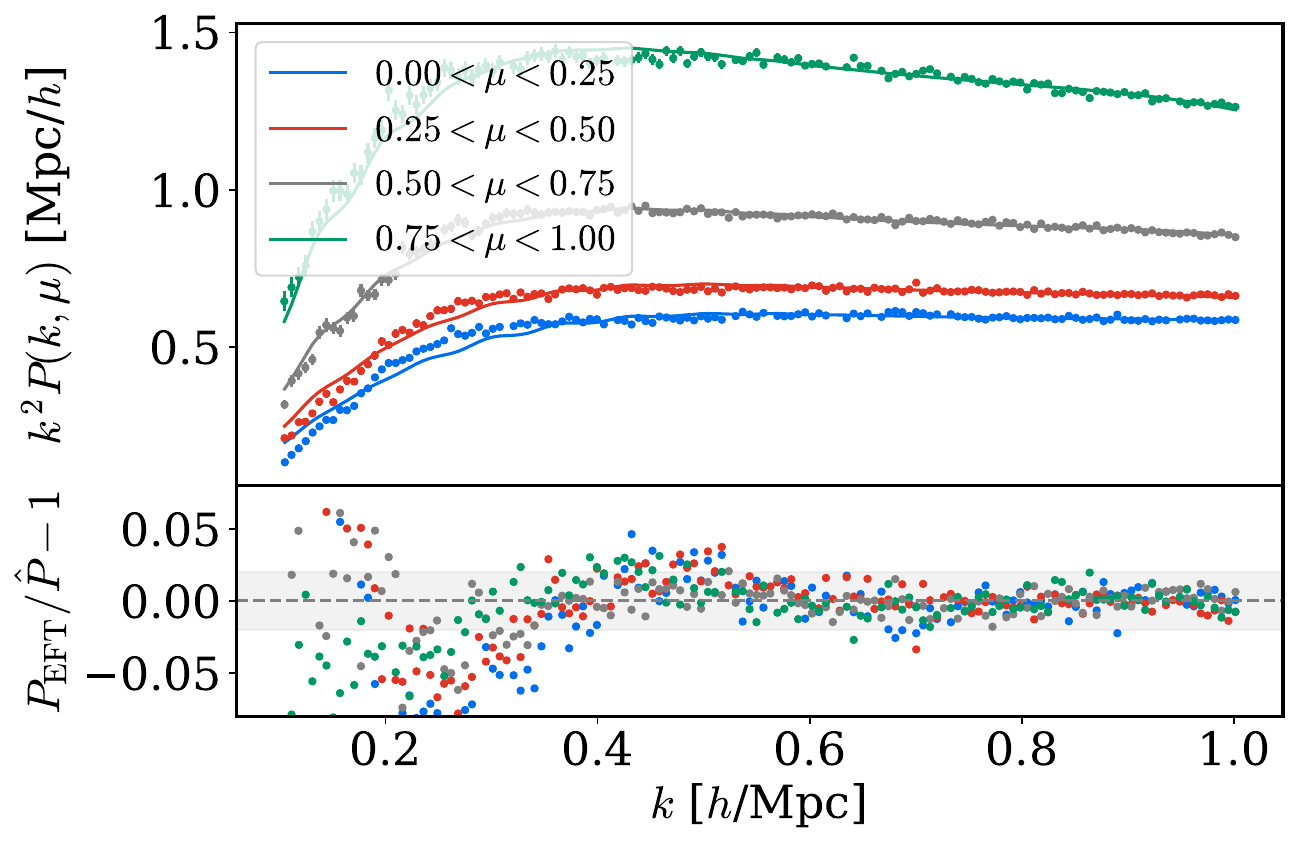}\hfill
    \includegraphics[width=0.32\linewidth]{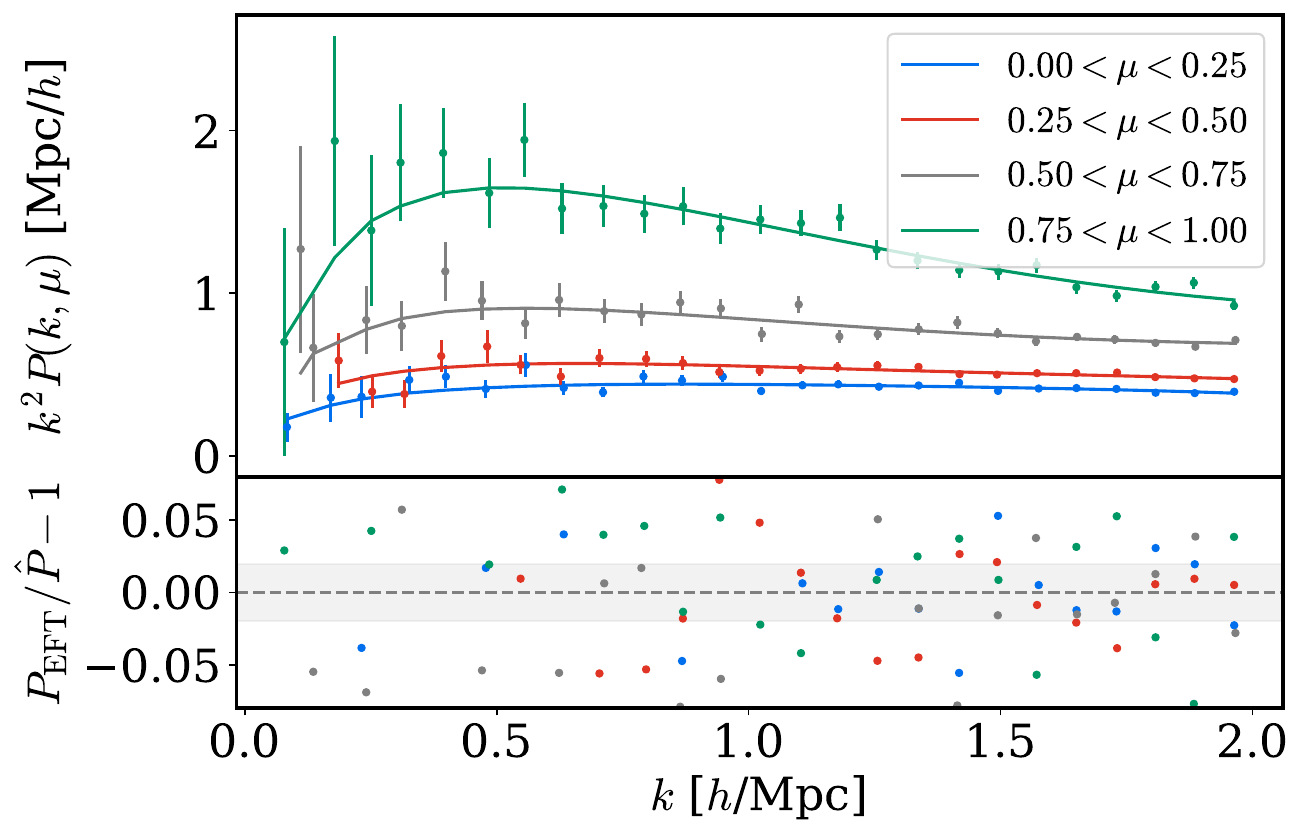}\hfill
    \caption{Comparison of the quality of the EFT fits to the 3D power spectra of the three simulations. \textit{Left panel:} Low-resolution box, \textit{Middle panel:} Reconstructed box, \textit{Right panel:} high-resolution input box as a function of angular bin $\mu$. In each bottom panel we show the residuals and to guide the eye with a 2\% error band shown in gray. We use a maximum wavenumber of $\kmax=1 \hMpcinv$ for the large simulations and $\kmax=2 \hMpcinv$ for the small volume. }
    \label{fig:EFT_fits}
\end{figure}

In this section, we show the quality of the fits in figure \ref{fig:EFT_fits}. It is interesting to note that the fits to the low- and high-resolution volumes yield very precise results. However, the perturbative model captures the small-scale features very well (mirroring the contour plot in figure \ref{fig:corner-plot}) but fails at capturing the large scales. This comparison transparently shows the regime where the generative model fails. The corresponding bias parameters for the fits are tabulated below in Table \ref{tab:EFT_fits}. 

\begin{table*}[h]
\centering
\begin{tabular}{lccc}
\hline\hline
$b_{\mathcal{O}}$ & Low-resolution & Reconstructed & High-resolution \\
\hline
$b_1$           & $-0.2821^{+0.0046}_{-0.0033}$ & $-0.0982 \pm 0.0018$ & $-0.1997^{+0.0321}_{-0.0295}$ \\
$b_{\eta}$      & $\phantom{-}0.0034 \pm 0.0071$ & $\phantom{-}0.0446^{+0.0032}_{-0.0032}$ & $\phantom{-}0.1672^{+0.1032}_{-0.0721}$ \\
$b_2$           & $\phantom{-}0.0995^{+0.3467}_{-0.3114}$ & $-0.1495 \pm 0.1046$ & $-0.5521^{+0.8004}_{-1.0348}$ \\
$b_{\mathcal{G}_2}$       & $\phantom{-}0.2469^{+0.2549}_{-0.2295}$ & $-0.1666^{+0.0989}_{-0.0911}$ & $-0.4863^{+0.4547}_{-0.6387}$ \\
$b_{\eta^2}$    & $\phantom{-}0.0210^{+0.1745}_{-0.1352}$ & $-0.7503^{+0.1028}_{-0.1286}$ & $\phantom{-}0.0707^{+0.8101}_{-0.6666}$ \\
$b_{\delta\eta}$& $-0.1179^{+0.7233}_{-0.2658}$ & $-0.7170^{+0.1977}_{-0.1793}$ & $-0.6516^{+0.8986}_{-0.5490}$ \\
$b_{(KK)_{\parallel}}$ & $-1.2576^{+0.9062}_{-0.4430}$ & $\phantom{-}0.5863 \pm 0.2429$ & $\phantom{-}0.4038 \pm 1.2596$ \\
$b_{\Pi^{[2]}_\parallel}$ & $-0.1151^{+0.4858}_{-0.1490}$ & $-0.5563^{+0.1044}_{-0.0768}$ & $-0.3691^{+0.5898}_{-0.4020}$ \\
\hline
$b_{\Pi^{[3]}_\parallel}$ & $\phantom{-}1.0830 \pm 0.1186$ & $\phantom{-}1.8838 \pm 0.2614$ & $\phantom{-}1.6372 \pm 0.6625$ \\
$b_{\delta\Pi^{[2]}_\parallel}$ & $-1.3370 \pm 0.2273$ & $\phantom{-}2.4556 \pm 0.4988$ & $-0.6425 \pm 0.7162$ \\
$b_{(K\Pi^{[2]})_\parallel}$ & $-1.2027 \pm 0.3184$ & $-1.7333 \pm 0.7309$ & $-0.6670 \pm 1.3111$ \\
$b_{\eta\Pi^{[2]}_\parallel}$ & $\phantom{-}0.3929 \pm 0.4059$ & $-0.1171 \pm 0.8765$ & $\phantom{-}0.1663 \pm 1.3458$ \\
$P_{\rm shot}$  & $\phantom{-}0.2043 \pm 0.3899$ & $\phantom{-}1.9225 \pm 0.2716$ & $\phantom{-}0.5650 \pm 0.3600$ \\
$a_0$           & $\phantom{-}0.1183 \pm 0.9992$ & $-1.6936 \pm 0.6955$ & $-0.3494 \pm 0.3167$ \\
$a_2$           & $\phantom{-}0.5170 \pm 0.8914$ & $-2.5761 \pm 0.9223$ & $\phantom{-}0.3657 \pm 0.4438$ \\
$c_0$           & $-0.0925 \pm 0.0922$ & $-1.8439 \pm 0.1859$ & $-0.2987 \pm 0.1509$ \\
$c_2$         & $\phantom{-}0.0088 \pm 0.0344$ & $\phantom{-}2.8017 \pm 0.0673$ & $\phantom{-}0.2454 \pm 0.1074$ \\
$c_4$         & $-0.0451 \pm 0.0134$ & $-1.9937 \pm 0.0244$ & $-0.1394 \pm 0.0462$ \\
\hline
\end{tabular}
\label{tab:EFT_fits}
\caption{Mean best-fit values for the one-loop EFT parameters obtained from the three simulations. The corresponding contours are shown in figure \ref{fig:corner-plot}. The default fit is performed with $\kmax=1\hMpcinv$ ($\kmax=2\hMpcinv$) for the large (small) boxes. The counterterms are divided by $(\hMpcinv)^2$ and we analytically marginalize over the parameters shown in the  bottom part of the table. Note that their posteriors are recovered \textit{a posteriori} from the chains.}
\end{table*}

\end{document}